# Electrical manipulation of skyrmions in a chiral magnet

Weiwei Wang[1,2], Dongsheng Song[1*], Wensen Wei[2], Pengfei Nan[1], Shilei Zhang[3], Binghui Ge[1], Mingliang Tian[2,4], Jiadong Zang[5,6*], and Haifeng Du[1,2,*]

[1]Institutes of Physical Science and Information Technology, Anhui University, Hefei 230601, China

[2]Anhui Province Key Laboratory of Condensed Matter Physics at Extreme Conditions, High Magnetic Field Laboratory, HFIPS, Anhui, Chinese Academy of Sciences, and University of Science and Technology of China, Hefei, 230031, China

[3]School of Physical Science and Technology, ShanghaiTech University, Shanghai 201210, China

[4]School of Physics and Materials Science, Anhui University, Hefei, 230601, China

[5]Department of Physics and Astronomy, University of New Hampshire, Durham, New Hampshire 03824, USA

[6]Materials Science Program, University of New Hampshire, Durham, New Hampshire 03824, USA

*Corresponding author:

dsong@ahu.edu.cn, Jiadong.Zang@unh.edu and duhf@hmfl.ac.cn

**Abstract**

Writing, erasing and computing are three fundamental operations required by any working electronic devices. Magnetic skyrmions could be basic bits in promising in emerging topological spintronic devices. In particular, skyrmions in chiral magnets have outstanding properties like compact texture, uniform size and high mobility. However, creating, deleting and driving isolated skyrmions, as prototypes of aforementioned basic operations, have been grand challenge in chiral magnets ever since the discovery of skyrmions, and achieving all these three operations in a single device is even harder. Here, by engineering chiral magnet $Co_8Zn_{10}Mn_2$ into the customized micro-devices for *in-situ* Lorentz transmission electron microscopy observations, we implement these three operations of skyrmions using nanosecond current pulses with a low a current density about $10^{10}$ A/m$^2$ at room temperature. A notched structure can create or delete magnetic skyrmions depending on the direction and magnitude of current pulses. We further show that the magnetic skyrmions can be deterministically shifted step-by-step by current pulses, allowing the establishment of the universal current-velocity relationship. These experimental results have immediate significance towards the skyrmion-based memory or logic devices.

The skyrmion lattice, a new magneto-crystalline order composed of topologically stable nanometer-sized magnetic whirls, was discovered in a chiral magnet MnSi a decade ago[1,2]. One year later, subsequent experiments showed that the skyrmion lattice can be driven to motion using direct currents (DC) at an ultra-low current density[3,4]. Such prominent feature together with its topological stability and small size opens the door to skyrmion-based spintronic devices[5], such as racetrack memory[6], logic devices[7], and neuromorphic computation[8,9]. However, from the technological point of view, the controllable creation, deletion and motion of isolated skyrmions[10–12], rather than skyrmion lattices, using nanosecond current pulses is of critical significance in practical applications[11,13,14]. These operations of the Néel skyrmions induced by the interfacial Dzyaloshinskii-Moriya interaction[15] (DMI) have been demonstrated in heterostructures using the spin-orbit torque (SOT)[13,14,16–20]. However, the same operations of skyrmions using spin-transfer-torque (STT) in bulk chiral magnets progress very slowly in the past decade. In contrast to the magnetic thin films, the main challenge is to fabricate high-quality microdevices from bulk sample for *in-situ* magnetic imaging. Very few experiments of electrical manipulation of isolated skyrmions[12,21] have been reported, but long and wide current pulses (milliseconds) had to be used.

Recently we have developed an efficient method of fabricating electrical micro-devices with arbitrary geometries used for *in-situ* Lorentz transmission electron microscopy (TEM). The focus ion beam (FIB) system was used, as shown in Supplementary Video 1. Based on this, the current-induced motion of a new type of multi-*Q* skyrmionic texture, termed as skyrmion bundle[22], has been demonstrated in a chiral magnet FeGe at the low temperature of $T \sim 95$ K. Here, we present systematic experiments on the STT-induced creation, deletion and motion of magnetic skyrmions in chiral magnet $Co_8Zn_{10}Mn_2$[23] at room temperature. Under a relatively low current density, the creation and deletion of isolated skyrmion have been realized via a geometrical notch at the device edge. Moreover, the universal current-velocity relation of skyrmions motions is established and a combined operation of skyrmions creation, motion and deletion is implemented.

## Skyrmions Creation and Deletion with a Notch

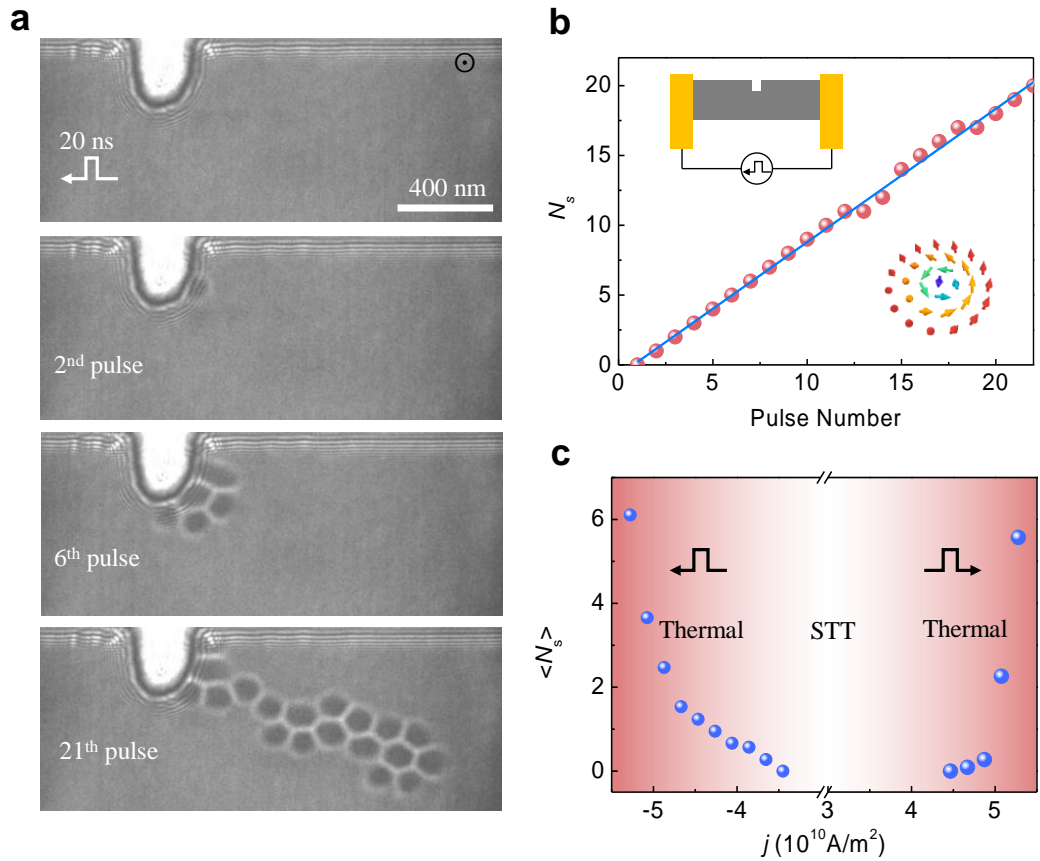

**Fig. 1 | Skyrmion creation with a notch. a**, A sequence of Lorentz TEM images of the skyrmion creation process after applying designated numbers of current pulses. The number of created skyrmion after the 2$^{nd}$, 6$^{th}$ and 21$^{th}$ pulse are 1, 5 and 20 respectively. The current density is $-4.26 \times 10^{10}$ A/m$^2$. The depth and width of the notch are 280 nm and 190 nm, respectively.

**b**, The number of created skyrmions as a function of pulse numbers. Inset: Schematic plots of the experimental setup (upper left) and a magnetic skyrmion (lower right). **c**, The average numbers of created skyrmions per current pulse as a function of current density. The creation rate is asymmetric with respect to current direction especially in the STT-dominated region. The magnetic field is $B = 70$ mT and pulse width 20 ns. The scale bar in **a** is 400 nm.

We first demonstrate the current induced skyrmion creation using a notch at the sample boundary[24]. The fabricated micro-device is composed of two Pt electrodes and a thin lamella with a thickness of ~150 nm (see Methods and Supplementary Fig. S1). A $190 \times 280$ nm² notch is specifically designed to serve as a nucleation seed for creating skyrmions using current. The notch width of ~190 nm is comparable to the period of spin helix ($L \sim 114$ nm)[25], much smaller than that reported in a recent FeGe-based device[12], making the creation of single skyrmion possible.

Fig.1**a** shows the snapshots of skyrmion creation process after applying a sequence of current pulses with the width of 20 ns and current density of $-4.26 \times 10^{10}$ A/m² in $x$ direction (see the details in Supplementary Video 2). Initially, the sample is in the conical state under the external field $B$ of 70 mT. After applying two pulses, one skyrmion with a topological charge of $Q = -1$ is created. Further application of current pulses continuously creates skyrmions one-by-one till the end of 12$^{th}$ pulse (Fig.1**b**). After that, it occasionally happens that no skyrmion is created under a few applications of current pulses. Nevertheless, the linear relationship between the number of created skyrmion and applied pulses is well identified (Fig. 1**b**). At last, a skyrmion cluster composed of 19 skyrmions is created after the 21$^{th}$ pulse. It stretches into a ribbon-like shape due to the skyrmion Hall effect[16,19,26].

The effect of the current pulse magnitude and direction on the skyrmion creation is summarized in Fig. 1c, where the parameter <$N_s$> represents the average number of created skyrmions per current pulse. Under a negative current pulse, the threshold current density for the skyrmion creation is approximately $-3.4 \times 10^{10}$ A/m², which is one order of magnitude smaller than the theoretical estimation[24]. On the contrary, a positive current with the same current density failed to create skyrmions (Supplementary Fig. S2), indicating the asymmetric STT effects with regard to the current direction[24]. This asymmetry of STT-induced skyrmion creation originates from the breakdown of reflection symmetry due to the unique direction of the spin precession region[24]. Consequently, skyrmions with $Q = +1$ could only be created with a positive current and a negative field (Supplementary Fig. S2). The thermal effect starts to dominate the creation process when the current density exceeds $5 \times 10^{10}$ A/m². After that, the Joule heating becomes increasingly prominent. The ultrafast field-warming beyond the $T_c$ and then field-cooling process results in the creation of skyrmions[27,28]. As a result, the unidirectionality is weakened (Supplementary Fig. S3 and Supplementary Video 3) and <$N_s$> rapidly increases (Fig. 1c).

The unidirectionality of skyrmion creation with the notch allows us to delete the skyrmion by making use of its inverse process. Fig. 2a shows the representative Lorentz-TEM images of a $N_s = 15$ skyrmion cluster after successive applications of negative current pulses under $j \sim 4.06 \times 10^{10}$ A m⁻² and $B \sim 70$ mT (see the details in Supplementary Video 4). Once the positive current pulses are applied, the number of skyrmions $N_s$ decrease quickly with the increase of pulse number. Finally, only one skyrmion is left. It is attached to the edge owing to the attractive interaction between skyrmion and the edge[29], which is different from the two-dimensional system where the boundary twist owing to DMI induces a repulsive potential to skyrmions.

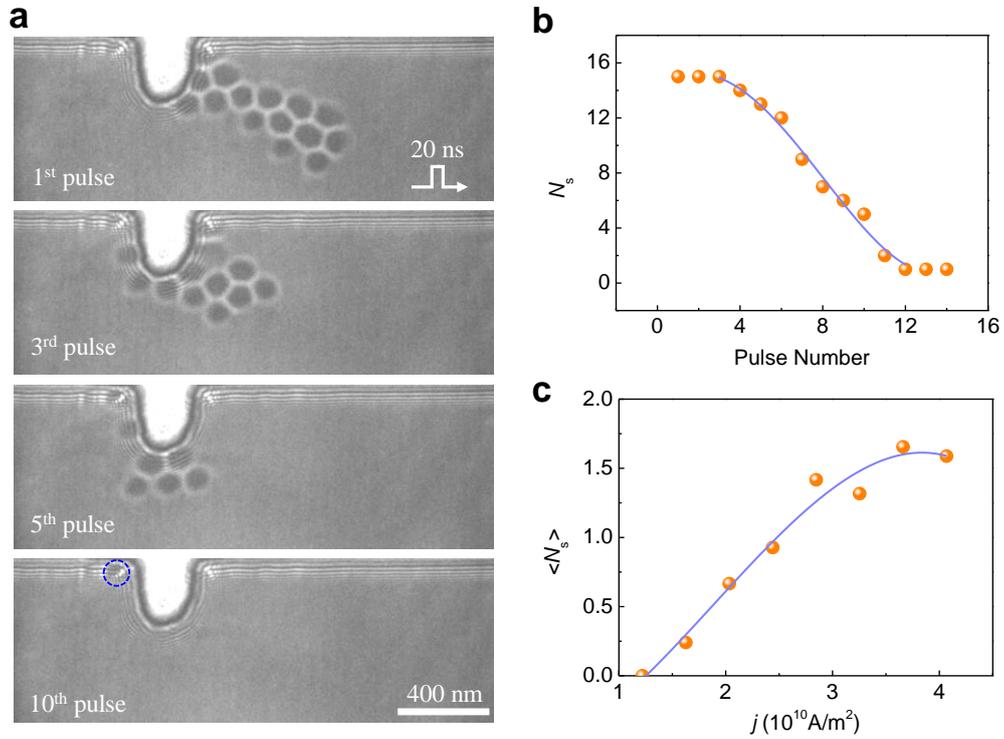

**Fig. 2 | Skyrmion annihilation with a notch. a**, A skyrmion cluster is pushed towards the notch using current pulses and absorbed by the notch gradually. The snapshots were taken under the defocus of 1 mm. The external field B = 70 mT, the pulse width is 20 ns and current density is $4.06 \times 10^{10}$ A/m². **b**, The number of skyrmions $N_s$ decreases as the current pulses applied to the system. **c**, The average number of erased skyrmion per pulse as a function of current density.

The number of remaining skyrmions as a function of current pulses is shown in Fig. 2**b**. The average number of deleted skyrmions per pulse depends on the strength of current density, as shown in Fig. 2**c**. The deletion rate increases with the current density (Supplementary Fig. S4) and reaches its maximum at $j \sim 3.65 \times 10^{10}$ A/m². Note that the threshold current density required to delete skyrmions is much smaller than that to create skyrmions although it appears to be the inverse process of the latter. It can be attributed to the asymmetric energy landscape between the conical state and the skyrmion state. The energy barrier of skyrmion deletion is smaller than that of skyrmion creation[30]. In principle, a flat edge should also be able to absorb skyrmions due to the inevitable skyrmion Hall effect under large current density[16]. However, it did not occur even at $j \sim 4.06 \times 10^{10}$ A/m² (Supplementary Fig. S5), indicating the crucial role of the rectangular notch.

**Skyrmions motion by current**

We now turn to the motion of skyrmions driven by STT. The universal current-velocity relation of skyrmion dynamics under STT has already been theoretically addressed[31]. The longitudinal velocity of skyrmion is derived as $v_x \approx -bj$ under the electrical current, where $b$ is a constant and $j$ is the current density (see Supplementary Note I). The experimental results of the nanosecond-pulse driven skyrmions motion are summarized in Fig. 3 with varied skyrmion numbers. For $Q = -1$, the skyrmion moves along the $+x$ direction under a negative current with pulse width of 80 ns and $j \sim -2 \times 10^{10}$ A/m² (Supplementary Video 5). The transverse motion along the $+y$ direction, i.e., skyrmion Hall effect (quantified as $\tan\theta_h = v_y/v_x$), is clearly observed, as depicted in Fig. 3**a**. The trajectory (Fig. 3**c**) shows approximately linear behavior as predicted in theory. The deviation from the linearity can be reasonably explained by the pinning effect and thermal fluctuation. The longitudinal velocity $v_x$ is always antiparallel to the current flow while the transverse velocity $v_y$ is related to the topological charge $Q$ (Supplementary Note I). Therefore, the magnetic skyrmion moves in the opposite direction when a positive current is applied and a reversal sign of $Q$ only changes the direction of velocity $v_y$ (Supplementary Fig. S6).

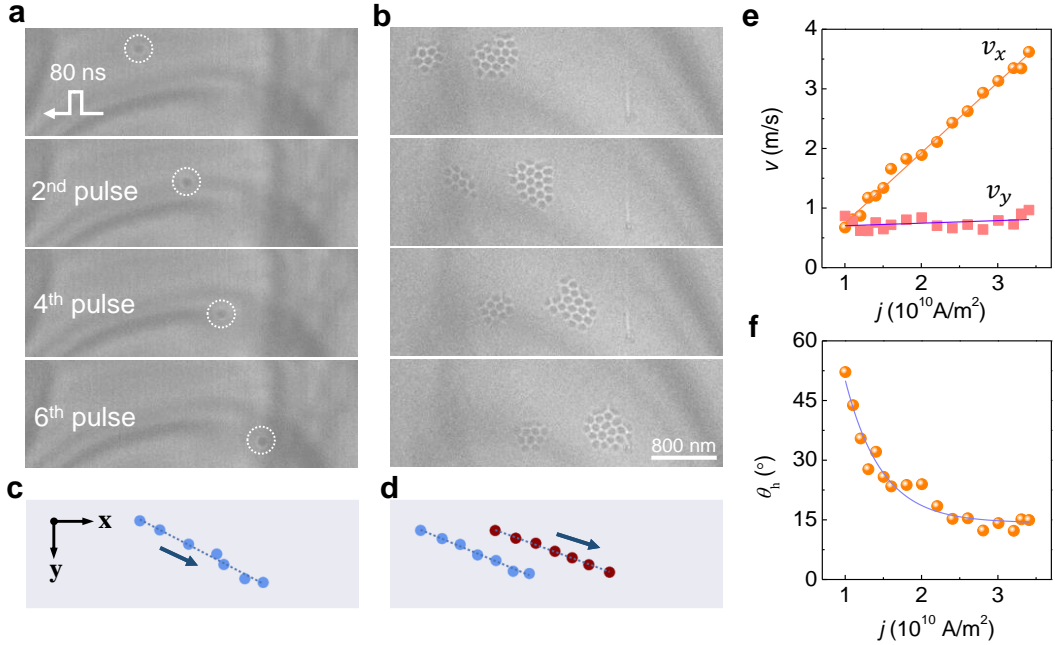

**Fig. 3 | Spin transfer torque induced single skyrmion and skyrmion clusters motion.** The sequences of images in **a** and **b** show the positions of skyrmions after applying the negative current pulses where the duration of each pulse is 80 ns and the current density is $-2 \times 10^{10}$ A/m². **a**, A single skyrmion with $Q = -1$. **b**, Two skyrmion clusters with $Q = -11$ and -21. The corresponding trajectories of the skyrmion center are shown in (**c** and **d**) and fit well to straight lines. The negative pulses lead to positive displacements in +$x$ direction which are independent on the skyrmion number. The nonzero transverse motion in the +$y$ direction is characterized by the skyrmion Hall angle $\theta_h$ which depends on the skyrmion number and is unrelated to the current direction. The amplitude of the external field $B$ is 117 mT. **e**, The skyrmion velocity as a function of the current density. The $x$-component velocity scales linearly with the current density. **f**, The skyrmion Hall angle $\theta_h$ as a function of current density.

The current-driven skyrmions motion can also be observed in skyrmion cluster states (Supplementary Video 6). Fig.3**b** show the collective motions of two skyrmion clusters with $N_s = -11$ and $N_s = -21$, respectively. The distance between two clusters remains constant during the motion. Both the velocity and skyrmion Hall angle are similar to those of the single skyrmion. However, the trajectories' deviation from straight lines are significantly suppressed with the increased skyrmions number in the cluster states (Supplementary Fig. S7). Interestingly, the skyrmion clusters can even steadily pass through a defect without noticeable deformation (Fig. 2**b** and Supplementary Video 6).

Based on the trajectories of magnetic skyrmions under varied current densities, the current-density-dependent skyrmion velocities are summarized in Fig. 3**e**. To minimize the uncertainty, skyrmion clusters with the number of $N_s \sim 20$ is selected therein. The predicted linear relationship[32] between the skyrmion velocity and current density is obtained. Moreover, the estimated spin polarization of current for $Co_8Zn_{10}Mn_2$ is $P \sim 0.57$ (see Methods), which is two times larger than that for $FeGe$[12] ($P \sim 0.27$), resulting in a comparable efficiency (defined as $\varepsilon = v_x/j$) to the reported record using SOT mechanism[13,16]. Below a low critical current density $j_{c1} = 1.0 \times 10^{10}$ A/m², magnetic skyrmions are static. This critical current density is directly related to the pinning forces arising from the disorder or impurity[32]. Above the critical density $j_{c2} = 3.5 \times 10^{10}$ A/m², skyrmions are dynamically created and annihilated due to the combined effect of STT and the Joule heating by current pulses[27].

Fig. 3**f** depicts an inverse relation between the skyrmion Hall angle ($\theta_h$) and the current density. In a defect-free system, the skyrmion Hall angle $\theta_h$ should be constant. However, in real materials, the defects-induced pinning force will give rise to a transverse motion of magnetic skyrmions, yielding an extrinsic skyrmion Hall effect[32]

(Supplementary Note I and Supplementary Fig. S9). At low current density, a low drift velocity increases the scattering rate and thus results in a large skyrmion Hall angle (Supplementary Fig. S8 and Supplementary Video 7). At higher current density, the scattering rate decreases and the observed skyrmion Hall angle is close to the intrinsic value $\theta_h$, which is as low as ∼15º. Our results are quite different from previous studies in magnetic multilayers[16,33,34], where the skyrmion Hall angle shows a complicated relation with current density. The underlying reason is that the skyrmion Hall angle therein is particularly susceptible to the change of radius under magnetic field and the deformation of spin texture in the motion.

## Integrated operations of skyrmions

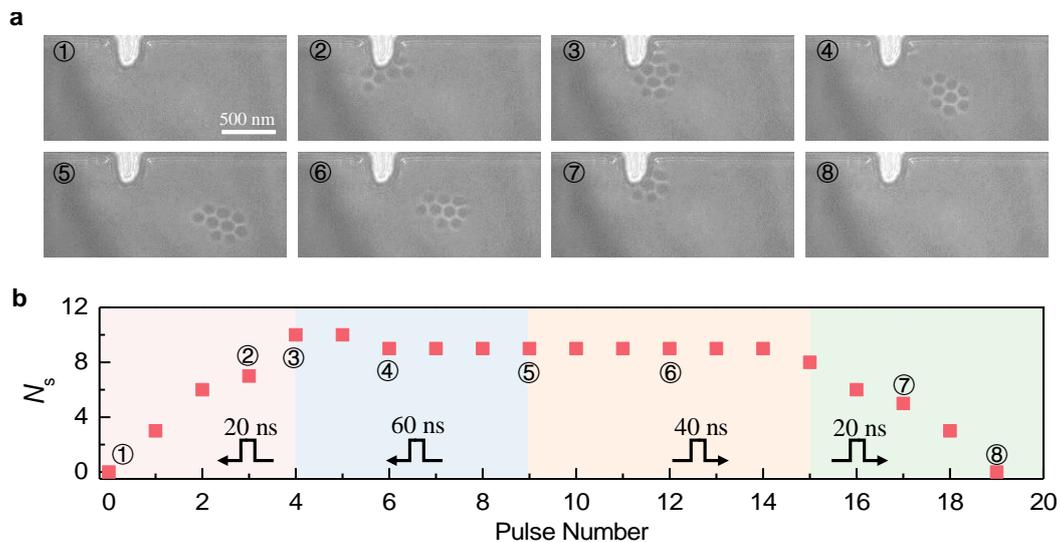

**Fig. 4 | Electrically manipulation of a skyrmion cluster: A combination of creation, motion and deletion. a,** The snapshots of the skyrmion cluster on different stages. The skyrmion cluster is created on the first stage (①②③) and moves forward on the second stage (③④⑤). The skyrmion then moves back on the third stage (⑥) and is finally deleted in the fourth stage (⑦⑧). **b,** The details of the current pulses and the number of skyrmions as a function of pulse

number are plotted. The current densities used on these four stages are $-4.86 \times 10^{10}$ A/m$^2$, $-2.03 \times 10^{10}$ A/m$^2$, $2.03 \times 10^{10}$ A/m$^2$ and $3.65 \times 10^{10}$ A/m$^2$, respectively.

At last, we demonstrate combination of all operations of creation, motion and deletion on a single device shown in Fig. 4 (see the details in Supplementary Video 8). On the first stage, a skyrmion cluster with 10 skyrmions is created with the pulse width of 20 ns and $j \sim -4.86 \times 10^{10}$ A/m$^2$ as shown in Fig. **4a**. On the second stage, the skyrmion cluster is displaced by 680 nm after five pulses with the pulse width of 60 ns and $j \sim -2.03 \times 10^{10}$ A/m$^2$. After that, the skyrmion cluster is pulled back with the pulse width of 60 ns at $j \sim 2.03 \times 10^{10}$ A/m$^2$. On the final stage, the skyrmion cluster is deleted eventually with the pulse width of 20 ns at $j = 3.65 \times 10^{10}$ A/m$^2$.

In summary, our work shows a proof-of-concept demonstration of necessary operations for skyrmion-based memory. This achievement of skyrmion creation, motion and deletion at room temperature in our experiments enables the chiral magnets as a unique platform for skyrmion-based spintronic devices. Additionally, chiral magnets allow the coexistence of other exotic particle-like magnetic objects such as bobbers[36] and hopfions[37,38], making the versatile spintronic devices[35] based on three dimensional spin textures possible.

# References

1. Mühlbauer, S. *et al.* Skyrmion Lattice in a Chiral Magnet. *Science* **323**, 915–919 (2009).

2. Yu, X. Z. *et al.* Real-space observation of a two-dimensional skyrmion crystal. *Nature* **465**, 901–904 (2010).

3. Jonietz, F. *et al.* Spin Transfer Torques in MnSi at Ultralow Current Densities. *Science* **330**, 1648–1651 (2010).

4. Schulz, T. *et al.* Emergent electrodynamics of skyrmions in a chiral magnet. *Nat. Phys.* **8**, 301–304 (2012).

5. Fert, A., Reyren, N. & Cros, V. Magnetic skyrmions: advances in physics and potential applications. *Nat. Rev. Mater.* **2**, 17031 (2017).

6. Fert, A., Cros, V. & Sampaio, J. Skyrmions on the track. *Nat. Nanotechnol.* **8**, 152–156 (2013).

7. Zhang, X., Ezawa, M. & Zhou, Y. Magnetic skyrmion logic gates: conversion, duplication and merging of skyrmions. *Sci. Rep.* **5**, 9400 (2015).

8. Song, K. M. *et al.* Skyrmion-based artificial synapses for neuromorphic computing. *Nat. Electron.* **3**, 148–155 (2020).

9. Grollier, J. *et al.* Neuromorphic spintronics. *Nat. Electron.* **3**, 360–370 (2020).

10. Romming, N. *et al.* Writing and Deleting Single Magnetic Skyrmions. *Science* **341**, 636–639 (2013).

11. Sampaio, J., Cros, V., Rohart, S., Thiaville, A. & Fert, A. Nucleation, stability and current-induced motion of isolated magnetic skyrmions in nanostructures. *Nat. Nanotechnol.* **8**, 839–844 (2013).

12. Yu, X. Z. *et al.* Motion tracking of 80-nm-size skyrmions upon directional current injections. *Sci. Adv.* **6**, eaaz9744 (2020).

13. Woo, S. *et al.* Observation of room-temperature magnetic skyrmions and their current-driven dynamics in ultrathin metallic ferromagnets. *Nat. Mater.* **15**, 501–506 (2016).

14. Legrand, W. *et al.* Room-Temperature Current-Induced Generation and Motion of sub-100 nm Skyrmions. *Nano Lett.* **17**, 2703–2712 (2017).

15. Rohart, S. & Thiaville, A. Skyrmion confinement in ultrathin film nanostructures in the presence of Dzyaloshinskii-Moriya interaction. *Phys. Rev. B* **88**, 184422 (2013).

16. Jiang, W. *et al.* Direct observation of the skyrmion Hall effect. *Nat. Phys.* **13**, 162–169 (2017).

17. Jiang, W. *et al.* Blowing magnetic skyrmion bubbles. *Science* **349**, 283–286 (2015).

18. Büttner, F. *et al.* Field-free deterministic ultrafast creation of magnetic skyrmions by spin–orbit torques. *Nat. Nanotechnol.* **12**, 1040–1044 (2017).

19. Litzius, K. *et al.* Skyrmion Hall effect revealed by direct time-resolved X-ray microscopy. *Nat. Phys.* **13**, 170–175 (2017).

20. Yu, G. *et al.* Room-Temperature Skyrmion Shift Device for Memory Application. *Nano Lett.* **17**, 261–268 (2017).

21. Yu, X. *et al.* Current-Induced Nucleation and Annihilation of Magnetic Skyrmions at Room Temperature in a Chiral Magnet. *Adv Mater* 6 (2017).

22. Tang, J. *et al.* Magnetic Skyrmion Bundles and Their Current-Driven Dynamics.
23. Tokunaga, Y. *et al.* A new class of chiral materials hosting magnetic skyrmions beyond room temperature. *Nat. Commun.* **6**, 7638 (2015).
24. Iwasaki, J., Mochizuki, M. & Nagaosa, N. Current-induced skyrmion dynamics in constricted geometries. *Nat. Nanotechnol.* **8**, 742–747 (2013).
25. Yu, X. Z. *et al.* Transformation between meron and skyrmion topological spin textures in a chiral magnet. *Nature* **564**, 95–98 (2018).
26. Zang, J., Mostovoy, M., Han, J. H. & Nagaosa, N. Dynamics of Skyrmion Crystals in Metallic Thin Films. *Phys. Rev. Lett.* **107**, 136804 (2011).
27. Zhao, X., Wang, S., Wang, C. & Che, R. Thermal effects on current-related skyrmion formation in a nanobelt. *Appl. Phys. Lett.* **112**, 212403 (2018).
28. Lemesh, I. *et al.* Current-Induced Skyrmion Generation through Morphological Thermal Transitions in Chiral Ferromagnetic Heterostructures. *Adv. Mater.* **30**, 1805461 (2018).
29. Du, H. *et al.* Interaction of Individual Skyrmions in a Nanostructured Cubic Chiral Magnet. *Phys. Rev. Lett.* **120**, 197203 (2018).
30. Zhang, X. *et al.* Skyrmion-skyrmion and skyrmion-edge repulsions in skyrmion-based racetrack memory. *Sci. Rep.* **5**, 7643 (2015).
31. Everschor, K. *et al.* Rotating skyrmion lattices by spin torques and field or temperature gradients. *Phys. Rev. B* **86**, 054432 (2012).
32. Iwasaki, J., Mochizuki, M. & Nagaosa, N. Universal current-velocity relation of skyrmion motion in chiral magnets. *Nat. Commun.* **4**, 1463 (2013).
33. Zeissler, K. *et al.* Diameter-independent skyrmion Hall angle observed in chiral magnetic multilayers. *Nat. Commun.* **11**, 428 (2020).
34. Woo, S. *et al.* Spin-orbit torque-driven skyrmion dynamics revealed by time-resolved X-ray microscopy. *Nat. Commun.* **8**, 15573 (2017).
35. Fernandez-Pacheco, A. *et al.* Three-dimensional nanomagnetism. *Nat. Commun.* **8**, 15756 (2017).
36. Zheng, F. *et al.* Experimental observation of chiral magnetic bobbers in B20-type FeGe. *Nat. Nanotechnol.* **13**, 451–455 (2018).
37. Wang, X. S., Qaiumzadeh, A. & Brataas, A. Current-Driven Dynamics of Magnetic Hopfions. *Phys. Rev. Lett.* **123**, 147203 (2019).
38. Liu, Y., Hou, W., Han, X. & Zang, J. Three-Dimensional Dynamics of a Magnetic Hopfion Driven by Spin Transfer Torque. *Phys. Rev. Lett.* **124**, 127204 (2020).

## Methods

### Preparation of $Co_8Zn_{10}Mn_2$ crystals

Polycrystalline sample of $Co_8Zn_{10}Mn_2$ crystals were synthesized by high-temperature reaction method. Stoichiometric amounts of cobalt (Alfa Aesar, purity > 99.9%), zinc

(Alfa Aesar, purity > 99.99%), and manganese (Alfa Aesar, purity > 99.95%) were loaded into a pure quartz tube and sealed under vacuum, heated to 1273 K for 24 h, followed by a slow cooling down to 1198 K, and then kept at this temperature for more than 3 days. After that, the tube was quenched into cold water. Finally, a ball-shaped $Co_8Zn_{10}Mn_2$ alloy with metallic luster was obtained.

## Fabrication of $Co_8Zn_{10}Mn_2$ micro-devices

The $Co_8Zn_{10}Mn_2$ micro-devices that suitable for TEM observation are fabricated from a polycrystal $Co_8Zn_{10}Mn_2$ alloy using the FIB-SEM dual-beam system (Helios NanoLab 600i; FEI) equipped with GIS, and Omniprobe 200+ micromanipulator. The detailed procedures can be found at Supplementary Fig. S1 and Video 1.

## Estimation of effective spin polarization for $Co_8Zn_{10}Mn_2$ and FeGe.

The effective spin polarization of current is established using the relation between the measured skyrmion velocity and the current density. Using the fitted slope $b = 1.18 \times 10^{-10}$ m$^3$/(A·s) and the estimated $M_s = 2.78 \times 10^5$ A/m for $Co_8Zn_{10}Mn_2$ at 300K, the spin polarization of $Co_8Zn_{10}Mn_2$ is estimated with P $\approx 2eM_s b/g\mu_B \approx$ 0.57. Similarly, the effective polarization for FeGe can be established as $P \sim 0.27$ using $M_s = 3.84 \times 10^5$ A/m and $b = 4 \times 10^{-11}$ m$^3$/(A·s) which is extracted from the Fig.3a in Ref.[12].

## Data availability

The data that support the plots provided in this paper and other finding of this study are available from the corresponding author upon reasonable request.

## Acknowledgments

H. D. acknowledges the financial support from the National Key R&D Program of China, Grant No. 2017YFA0303201; the Key Research Program of Frontier Sciences, CAS, Grant No. QYZDB-SSW-SLH009; the Key Research Program of the Chinese Academy of Sciences, Grant No. KJZD-SW-M01; the Strategic Priority Research Program of Chinese Academy of Sciences, Grant No. XDB33030100; and the Equipment Development Project of Chinese Academy of Sciences, Grant No. YJKYYQ20180012.

## Author contributions

H.D. supervised the project. J.Z. conceived the theory. W.W., D.S., P. Nan and H.D. conceived the experiments. W-S.W. synthesized $Co_8Zn_{10}Mn_2$ crystals. W.W., D.S., H.D. and J.Z. prepared the manuscript. All authors discussed the results and contributed to the manuscript.

## Competing interests

The authors declare no competing interests.

## Additional information

Supplementary information is available for this paper.